# Magnetic nanorods for quantitative viscosity imaging using heterodyne holography

C. GENTNER[1], P. BERTO[1,2], J.-F. BERRET[3], S. REICHMAN[1], R. KUSZELEWICZ[1], G. TESSIER[1]

[1] Sorbonne Université, INSERM, CNRS, Institut de la Vision, 17 rue Moreau, F-75012, Paris, France

[2] Université Paris Cité, 45 Rue des saints Pères, F-75006 Paris, France

[3] Université Paris Cité, Laboratoire Matière et Systèmes Complexes, 10 rue A. Domon et L.Duquet F-75205 Paris Cedex 13 France

**Abstract**

Many processes in microfluidics and biology are driven or affected by viscosity. While several methods are able to measure this parameter globally (AFM, surface acoustic waves, DLS, …), very few can provide high resolution viscosity images, particularly in living cells. Here, we propose to use sub-micrometer magnetic rods to perform high resolution viscosity imaging. An external magnetic field forces the oscillation of superparamagnetic iron oxide nanorods. Under linearly polarized illumination, the rotation of these highly anisotropic optical scatterers induces a blinking which is analyzed by heterodyne holography. The spectral analysis of the rotation dynamics yields a regime transition frequency $f_t$ from which the local viscosity is deduced. Holography provides 3D image reconstruction and 3D superlocalization of the nanorods, which allows super-resolved viscosity measurements. Relying on the fast Brownian rotation of nanorods instead of their slower translation therefore allows faster measurements and, crucially, smaller effective voxels for viscosity determination. Viscosity imaging is demonstrated with a 0.5 $\mu m^3$ 3D resolution.

**Intro**

Viscosity-driven phenomena are ubiquitous in current biological research concerns[1]. For instance, viscosity has been established as a key factor in cancer cells identification[2], but also in embryonic development through the off-center positioning of the oocyte nucleus during meiosis[2,3]. While macroscopic and global viscosity measurement techniques are well established[4, 5, 6, 7, 8, 9], smaller scale mapping methods are needed to unveil cytoplasm and cytoskeleton structures and behaviors.

Viscosity essentially quantifies the response of a medium to mechanical actuation[10]. Its measurement therefore requires detectable transducers to i) induce and ii) quantify movements. Although invasive, nano- or microscopic probes are the only current way to produce microscopic viscosity mapping in a closed environment such as a cell[11]. External magnetic actuation is probably the most preeminent method to induce probe movements. In several microrheology experiments[12, 13, 14, 15, 16, 17, 18], maghemite rods (a polymorphic iron oxide[18, 19]) are rotated using various revolving magnetic field frequencies. The cut-off frequency at which rods cease to follow the actuation provides an accurate measurement of the viscosity. However, since rotation is detected using direct optical imaging, the rods must be resolved, and their sizes typically lie in the 10 µm range. While intracellular measurements have been reported, they can hardly be considered local since the probe size is comparable to that of most cells[18]. Optical tweezers are another tool to induce movement. They can drive particles with high precision, and allow measurement duration as short as tens of µs[20, 21, 22, 23]. Imaging can then be obtained using multiple optical tweezers and/or particle scanning.

However, to obtain full images of viscosity, multiple, freely moving probes using Brownian motion as an inherent actuation are potentially faster and simpler. Fluorescence-based probes are therefore widely used for microrheology[24] [25] [26]. Rather than translation, they rely on fluorescence lifetime imaging (FLIM) and on the rotation of the probe molecules, the latter occurring on shorter time scales than translation processes[27,14]. Since probes are permanently in translational motion, viscosity is averaged over the volume which the probe explores during the measurement. With a freely-diffusing probe, the best possible spatial resolution for a viscosity measurement is therefore given by $\sqrt{D_{lat}\tau}$, where $D_{lat}$ is the translational diffusion coefficient and $\tau$ is the measurement duration. Fluorescence-based methods relying on fast rotational movements offer an interesting way to reduce $\tau$ and preserve the locality of the measurements. However, in addition to possible photobleaching or phototoxicity issues, fluorescence-based viscosity measurements can hardly be used with single molecule resolution: low photon yields require longer acquisition times $\tau$, during which Brownian translation counteracts the accuracy of single molecule superlocalization.

Nanoparticle probes therefore appear as promising alternatives for viscosity measurements. In metallic particles, plasmonic resonances can be exploited to obtain relatively strong scattering cross sections. An illumination at the appropriate wavelength (plasmonic resonance) and power (to mitigate thermal effects) allows the detection of particles between a few tens and a few hundreds of nanometers within tens of milliseconds. Although such nanoparticles are 1 or 2 orders of magnitude larger than fluorescent molecules, they can actually lead to better spatial resolutions, owing to lower diffusion coefficients $D_{lat}$ and shorter detection times. As in the case of fluorescent molecules, relying on Brownian rotation rather than translation is faster and more effective. In the case of nanoparticles, a good option is to use asymmetric objects like nanorods. Optical scattering by a nanorod is characterized by an anisotropic radiation pattern[28]. As the rod rotates, this results in an orientation-dependent modulation of the scattered light[29], which has an optimal contrast when the rod is illuminated near the plasmonic resonance of one of its axes. Nanorods illuminated by a linearly polarized laser beam as they undergo Brownian rotation in a fluid therefore induce a blinking. A spectral analysis of this blinking can lead to local viscosity values, allowing a quantitative viscosity mapping[30]. However, the random nature of the rotation, and of the associated blinking, calls for a careful averaging and an analysis of the frequency spectrum of the optical fluctuations, which is time-consuming and hardly compatible with fast imaging.

In this paper, we report on a fluorescence-free method which takes advantage of the simplicity and efficiency of an external magnetic actuation to rotate superparamagnetic rods, but scales down the probe size by more than one order of magnitude. Since the size of these rods is of the same order as the diffraction limit, determining their orientation through direct imaging is difficult, and we chose to detect the variations in their effective scattering cross section as they rotate under a uniform polarized illumination. Using digital holography and a heterodyne lock-in scheme, the blinking signal of multiple or individual rotating probes is efficiently extracted from the background, from which we deduce quantitative, local and 3-dimensional values of the surrounding fluid viscosity. Viscosity measurements with a single 3D-tracked superparamagnetic rod is demonstrated, paving the way to super-resolved viscosity imaging.

1. **Principle**

When aiming at high spatial resolution, the size of the probes is a key issue. Since suitably small magnetic nanorods are not commercially available, sub-micrometer maghemite rods (length $L = 1020 \pm 240$ nm, diameter $D = 440 \pm 80$ nm, mean values and standard deviations measured by imaging 50 rods magnetically aligned in the image plane) were synthesized. As shown in Fig.1a, the orientation of these rods is driven by an external magnetic field modulated at a frequency $f_B$, superimposed to a perpendicular static field, thus creating an oscillating effective magnetic field $\vec{B}(f_B)$.

The optical detection of the rods is based on the blinking resulting from rod rotation as described previously for nonmagnetic objects[30]. The scattering diagram of a nanorod is essentially a torus aligned with the long rod axis, as schematically depicted in Fig.1a. When illuminated by a $785\ nm$ laser beam with a polarization

$\vec{E_O}$, the intensity and angular distribution of the electric field $\vec{E_S}$ scattered by a rod depend on its orientation with respect to $\vec{E_O}$. The portion of this light collected by a microscope objective is therefore modulated as the rod rotates. A Digital Holographic Microscope (DHM) operating in homodyne or heterodyne lock-in mode[31] selectively measures the amplitude of the collected light modulated at a chosen frequency $f$ (see section 2). This interferometric imaging method provides two main advantages. First, DHM exploits the product $\vec{E_S} \cdot \vec{E_R}$ of the electric field $\vec{E_S}$ scattered by the object (i.e. the light scattered by a nanorod) by a reference field $\vec{E_R}$. Although $E_S$ is weak in the case of nanorods with small scattering cross section, $E_R$ can be easily increased to obtain hologram intensities approaching the saturation level of the camera, thus providing an optimal, shot noise-limited detection of small nanoparticles[23, 24, 30, 31]. Second, the two-arm interferometric configuration is particularly suited to single- (homodyne) or dual- (heterodyne) frequency phase modulations. As such, it allows the study of modulated phenomena over a very broad frequency range (from mHz to tens of MHz)[32, 33, 34] [35], and thus gives access to high rotation/blinking frequencies.

As shown in Fig.1b, when monitoring a single nanorod, the presence of a magnetic field $\vec{B}$ modulated at $f_B = 10$ Hz induces an increase in the DHM signal at the frequency $f = f_B$ which clearly indicates that the rod rotates and blinks at the frequency $f_B$. However, as shown in the inset, the object does not disappear totally in the absence of modulation: Brownian motion randomly contributes to a component at $f$.

A simultaneous scanning of the magnetic field and analysis frequencies $f_B$ and $f$ yields a spectrum which displays two distinct frequency regimes separated by a transition at a critical frequency $f_t$, as shown in Fig.1c (see Supplementary):

- $f \ll f_t$: the magnetic field oscillation period is longer than the characteristic time taken by the magnetic moment to align. The rod is able to follow the rotation of the magnetic field, and the signal amplitude is then maximum.

- $f \gg f_t$: the magnetic field oscillation period is too short to allow the rod to align with it. The average orientation of the rod long axis is thus stuck close to the direction of the permanent magnetic field $\vec{B_0}$, and the amplitude of the modulated component of the DHM signal decreases to zero as $f_B$ increases.

This transition frequency $f_t$ is clearly related to viscosity $\eta$. Its measurement therefore gives access to $\eta$ through Eq. (1), which is obtained by considering the evolution of a superparamagnetic rod, i.e. a *macro-magnetic* moment $\vec{m} = \frac{\bar{\chi}V}{\mu_0}\vec{B}$ [18], in a field $\vec{B} = B_\sim cos(2\pi F_B t)\vec{u_z} + B_0\vec{u_x}$. (See Supplementary for derivation).

$$f_t = \frac{1}{2\pi\mu_0}\frac{\Delta\chi B^2}{\eta}h\left(\frac{L}{D}\right)$$
(1)

where $\eta$ is the medium dynamic viscosity, $h$ is a function of the rod length $L$ and diameter $D$ defined as $h\left(\frac{L}{D}\right) = \frac{L^2 N_\parallel + D^2 N_\perp}{2(L^2+D^2)}$ with spheroidal depolarization factors expressed in terms of the eccentricity $e = \sqrt{1-\left(\frac{D}{L}\right)^2}$ as $N_\parallel = \frac{1-e^2}{2e^3}\left(ln\left(\frac{1+e}{1-e}\right) - 2e\right)$, $N_\perp = \frac{1-N_\parallel}{2}$. $\Delta\chi = \frac{\chi^2(N_\perp - N_\parallel)}{(1+N_\parallel\chi)(1+N_\perp\chi)}$ is the magnetic susceptibility anisotropy factor for a rod composed of an assembly of spherical nanoparticles of massive susceptibility $\chi$ [36]. The factor $B^2$ in Eq. (1) corresponds to the squared modulus of the average magnetic field, $B^2 = B_0^2 + \frac{B_\sim^2}{2}$.

As shown in Supplementary, the measured spectrum can be modeled by a sigmoidal function, convoluted with a $sinc$ transfer function[30]. The inflexion point of this sigmoid indicates the transition frequency $f_t$. For

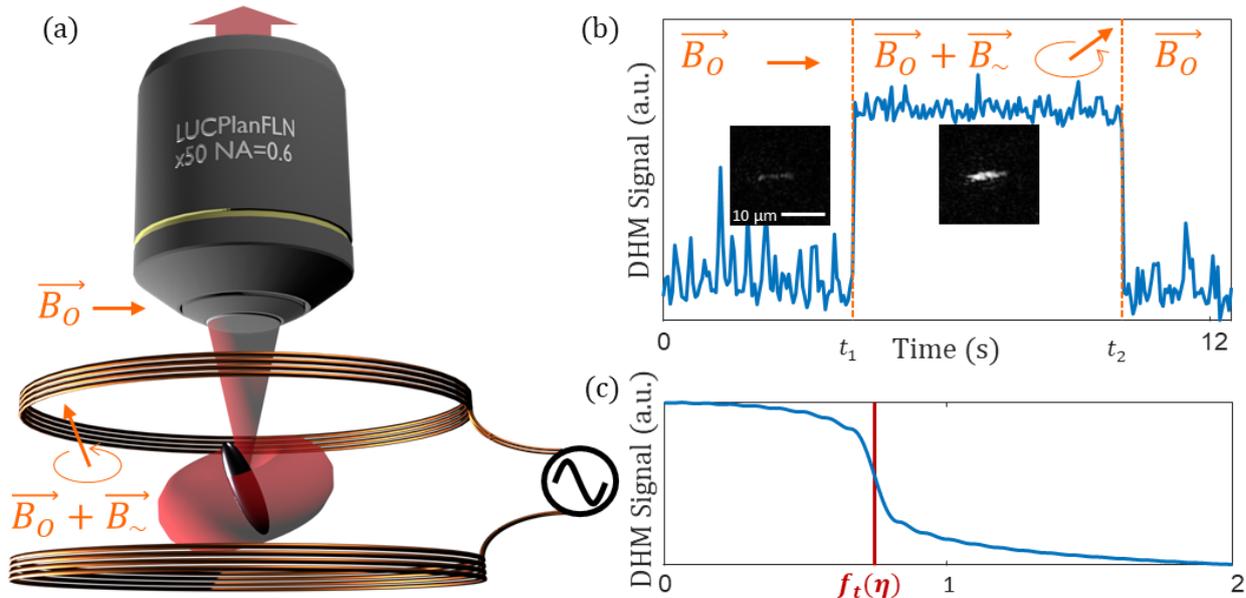

Figure 1 (a) Principle of the measurement: the orientation of a maghemite nanorod is driven by the superposition of a static magnetic field $\vec{B_0}$ (horizontal) and an oscillating field $\vec{B_\sim}$ (vertical) modulated at a frequency $f_B$. Part of the torus-shaped optical scattering diagram is collected by the microscope objective and sent to a heterodyne Digital Holography Microscope (DHM). (b) Reconstructed DHM images (insets) and average signals acquired at an analysis frequency $f = f_B = 10$ Hz. When $\vec{B} = \vec{B_0}$ ($\vec{B_\sim} = \vec{0}$ for $t \in [0, t_1[\cup[t_2, +\infty[)$, the rod is mostly static, aligned with $\vec{B_0}$, and the $f_B$ component of the DHM signal is low. When $\vec{B} = \vec{B_0} + \vec{B_\sim}$ ($t \in [t_1, t_2[$), the strong DHM signal at $f_B$ indicates that the rod oscillates and blinks at $f_B$. (c) Numerical model of the DHM signal, a sigmoid with a $sinc$ acquisition transfer function[30].

a rod of dimensions $L$ and $D$ in an oscillating magnetic field of amplitude $B$, a fit of this spectrum using Eq. (1) with $f_t$ as parameter can therefore provide a local value of the viscosity $\eta$ across the volume explored by the rod during the acquisition.

The expression of $f_t$ in Eq. (1) is similar to that of the more widely studied case of a rotating field[13, 16, 18], although the transition is broader: as illustrated in Supplementary, when the oscillating field frequency increases, the angular exploration range of the rod decreases smoothly from its maximum (lower than $\pi$) to 0 within the same half-plane; whereas a rotating field produces a sharp cutoff between a full $2\pi$ rotation regime and a regime with oscillation around a fixed (arbitrary) axis. Rotating fields are typically produced using four identical coils[12-14, 18]. Here, the dark field illumination through an oil-coupled prism strongly constrains the available space around the sample, and we chose a more compact oscillating field configuration.

Depending on the investigated medium and on the geometry of the rods, the frequency range of the transition frequency $f_t$ imposes the use of either homodyne (low frequency) or heterodyne (higher frequencies) holography, as detailed in the following sections.

## 2. Low frequencies - high viscosities: homodyne holography

The first experiment, shown in Fig.2, was carried out in two microfluidic chambers separated by a horizontal 300 μm PDMS wall. The first chamber (Fig.2b-top) contains a 27% glycerol aqueous solution ($\eta = 2.6$ mPa s). The second chamber contains water ($\eta = 1.0$ mPa s) for comparison. The $1020 \times 440$ nm maghemite nanorods were synthesized using a low weight concentration ($c = 0.02$ wt%) of initial nanoparticle-polymer solution using a synthesis protocol described in [37,38]. Individual nanorod detection requires a density well below the limit of 1 object per optical Point Spread Function (PSF), i.e. a few $10^6$ rods.μL$^{-1}$ with the X4, NA=0.13 objective we used. Here, the rods were diluted to $10^3$ μL$^{-1}$ in both chambers, 3 orders of magnitude below this limit. The rods were submitted to an oscillating magnetic field $\vec{B} = \vec{B_0} + \vec{B_\sim}$, with a modulation amplitude $B_\sim = 1.3 \pm 0.3$ mT, and a static component $B_0$ in the mT range.

The holographic detection setup is depicted in Fig.2a. To allow an intense illumination of the low-scattering-cross-section nanoparticles without blinding the camera, the beam illuminating the microfluidic chamber is sent through a glass prism coupled to the microfluidic chamber, and totally reflected at the glass-air interface. Only the light scattered by the nanorods is collected by a X4, NA=0.13 objective and sent to the camera. There, it interferes with a reference plane-wave coming from the same 785 nm, single longitudinal mode laser in an off-axis configuration, at an angle ~1°. In order to extract the component of the signal modulated at $f_B$, the camera is triggered at $f_{acq} = 4f_B$, in a 4-bucket integration scheme[34, 39] which filters out fluctuations occurring at other frequencies and selectively produces holograms of the $f_B$-modulated component. After a first Fourier transform, the +1 interferometric order is numerically extracted, propagated, and finally back Fourier-transformed to obtain the image of a chosen plane inside the microfluidic chamber[40].

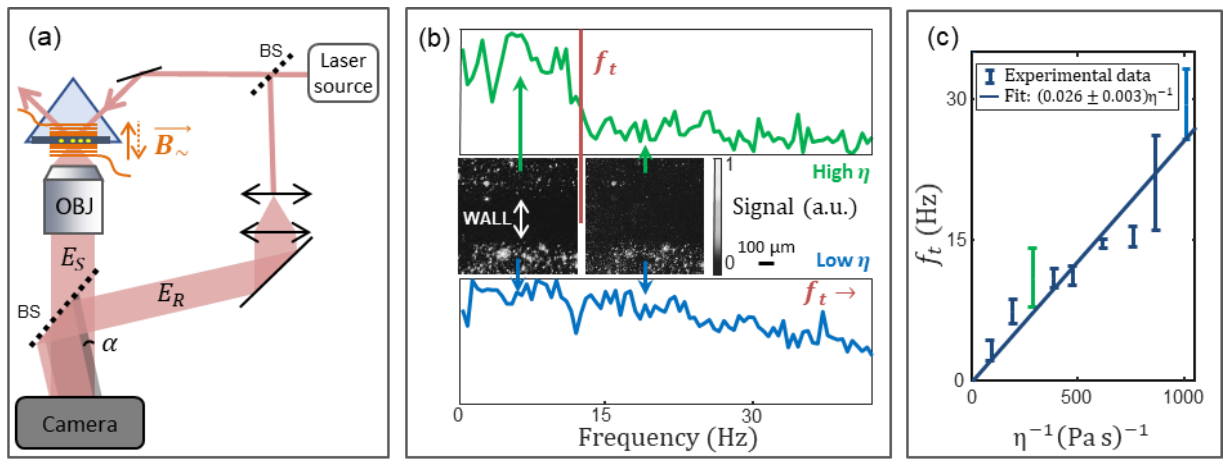

*Figure 2* (a) Holographic setup. A magnetic field oscillating at $f_B$ modulates the orientation and optical scattering of rods placed in a microfluidic chamber illuminated in total internal reflection. The off-axis holograms are sampled at $f_{acq} = 4f_B$, demodulated and reconstructed. (b) Images of two chambers containing water-glycerol mixtures of viscosity 2.6 mPa s (top) and 1.0 mPa s (bottom) separated by a horizontal wall, at frequencies $f_B = 7$ Hz and $f_B = 20$ Hz. Spectra of signal averaged in the high viscosity (top, green) and low viscosity regions (bottom, blue) exhibit different transition frequencies $f_t$, at 12 Hz for $\eta = 2.6$ mPa s and 35 Hz for $\eta = 1.0$ mPa s. (c) Transition frequency $f_t$ measured in solutions of 0 to 55% water-glycerol mixtures. Error bars: dispersion obtained on 10 different rods.

For each frequency $f$ in the range [0.5,42] Hz, series of 80 holograms were acquired, yielding 20 holograms after 4-bucket demodulation. Fig.2b shows images reconstructed from the average of these 20 holograms obtained at 7 and 20 Hz. Similar images and signals averaged in each compartment were acquired in 0.5 Hz steps. The values of $f_t$ were extracted from fits using the function derived in detail in the Supplementary section. Note that a simplified sigmoidal function was used here, omitting the convolution by the $sinc$ function for sake of computational speed.

In the high viscosity compartment ($\eta = 2.6$ mPa s, green curve in Fig.2b), a clear transition frequency $f_t = 12 \pm 1$ Hz is observed. With $\frac{L}{D} = 2.8 \pm 0.5$ and $\chi = 6.5$[18], Eq. (1) predicts $f_t = 12 \pm 7$ Hz for $B_0 = 0.9 \pm 0.3$ mT, in good agreement with the measured value. In the lower viscosity medium, $\eta = 1.0$ mPa s, we measure a transition at $f_t = 30 \pm 5$ Hz, which compares well with the expected value $f_t = 32 \pm 22$ Hz. Images acquired between these two values of $f_t$ (Fig.2b, center, $f = 20$ Hz) clearly display a viscosity contrast: while rods do not follow the rotation of the magnetic field in the high viscosity medium (top), they are rotating and appear bright in the low- region (bottom).

Similar spectra were acquired to measure $f_t$ in various water/glycerol mixtures, with glycerol concentrations in the [0%, 55%] range, corresponding to viscosities in the [1,10] mPa s range (deduced using [41]). The measured values $f_t(\eta^{-1})$ plotted in Fig.2c display a linear behavior, with a best fit slope of $0.026 \pm 0.003$ Pa. This is in good agreement with the value $f_t\eta = \frac{\Delta\chi B^2 h\left(\frac{L}{D}\right)}{2\pi\mu_0} = 0.03 \pm 0.01$ Pa predicted by Eq. (1) with the parameters used in our experiment, $\Delta\chi = 2.0 \pm 0.1, \frac{L}{D} = 2.8 \pm 0.5, B_0 \approx 1$ mT, $B_\sim = 1.3 \pm 0.2$ mT.

However, the accurate determination of the transition frequency requires measurements at frequencies $\frac{f_{acq}}{4}$ well above the value of $f_t$. With $f_t = \frac{\Delta\chi B^2 h\left(\frac{L}{D}\right)}{2\pi\eta\mu_0}$, this can be difficult to achieve with most cameras ($f_{acq}$ of a few tens of Hz), particularly at low viscosity values.

A possible workaround is to decrease the intensity of the modulated component of the magnetic field $B_\sim$. However, this approach breaks down when the magnetic driving force becomes of the same order as the random Brownian forces exerted on the particle. Before this, lowering $B_\sim$ clearly decreases the detection signal-to-noise ratio. If we choose an *arbitrary* limit of $10^3$ for the ratio between the magnetic energy $mB$ and the energy of Brownian fluctuations $k_BT$ which allows a satisfactory detection, i.e. $\frac{mB}{k_BT} > 10^3$, we can evaluate the minimal magnetic modulation $B$ which is required. Superparamagnetic rods have a macroscopic magnetic moment amplitude $m = \frac{\chi V}{1+N_\parallel\chi}\frac{B}{\mu_0}$ where $V$ is the rod volume. For our rod size and composition, the minimum magnetic field amplitude needed to overcome Brownian fluctuations is then $B = 3.5$ mT, which, from Eq. (1), corresponds to a transition frequency $f_t = 200$ Hz in water. In 4-bucket integration mode, its measurement therefore requires camera frame rates well above $4f_t = 800$ Hz. Cameras operating in the kHz range or above are available, but sub-ms integration times are hardly compatible with nano-object detection. For this reason, we have chosen to use heterodyne holography, which allows the measurement of high frequency phenomena using slow detectors[30, 31-35, 39, 40, 42].

### 3. High frequencies: heterodyne holography

Heterodyne detection is implemented by adding an acousto-optic modulator (AOM) in each holographic path, as illustrated in Fig.3a, to introduce a relative phase modulation at a chosen frequency. Since AOMs operate in the 80 MHz range, the object beam is modulated at 80 MHz, while the reference beam is modulated at $80$ MHz $-\Delta f$ to produce an optical phase beating at $\Delta f$. The 4-bucket integration described above is used to sample and demodulate holograms with a camera triggered at $f_{acq}$. By choosing the beating frequency so that $\Delta f = \frac{f_{acq}}{4} + f_i$, one can selectively isolate holograms of any modulation at $f_i$. Much like a multiplexed lock-in scheme, this process isolates a hologram of component modulated at frequency $f_0$ and filters out other fluctuations (including most of the Brownian noise in our case). Even with a slow camera, i.e. $f_{acq}$ of a few Hz, heterodyning provides high-SNR-holograms of phenomena modulated at frequencies $f_i$ up to tens of Hz. To illustrate the ability of this system to image quantitatively the viscosity, a test sample displaying a spatial viscosity gradient was prepared using a concentration $C_{Agar} = 0.5$ wt% Agar gel ($\eta \approx 5$ mPa s$^{[43]}$) and water ($\eta = 1.0$ mPa s). These media were put in contact to obtain a gradient of Agar concentration by diffusion at the interface. In order to visualize this gradient, 1% Alexa 488 fluorophores were added to the water phase. The fluorescence image shown in Fig.3c-left is therefore a direct image of the water gradient ($1 - C_{Agar}$), which is directly related to the viscosity. Both media were homogeneously seeded using the same maghemite nanorods ($L = 1020 \pm 240$ nm, $D = 440 \pm 80$ nm) at a concentration of $10^5 \mu L^{-1}$. With the microscope objective used here (×4 ON 0.13), this corresponds to about 100 rods per PSF, i.e. rods were not individually resolved.

In the experiment shown in Fig.3, the coils were powered using a 100 V – 200 W audio amplifier to produce a modulated field amplitude $B = 3.9$ mT. This stronger field modulation (as compared to $B = 1.3$ mT in the previous experiment) increases SNR, but also leads to higher transition frequencies $f_t$. Heterodyning, however, gives access to a broad frequency range. For each of the 27 measured frequencies $f_i$ (i.e. $i = 1$ to 27) ranging from 80 to 1000 Hz, 100 holograms were acquired at $f_{acq} = 10$ Hz, with a $\tau = 50$ ms integration time. After 4-bucket demodulation of the 100 measurements, the 25 resulting holograms were averaged and numerically propagated to the sample plane to produce 27 images of the dynamic blinking for each of the 27

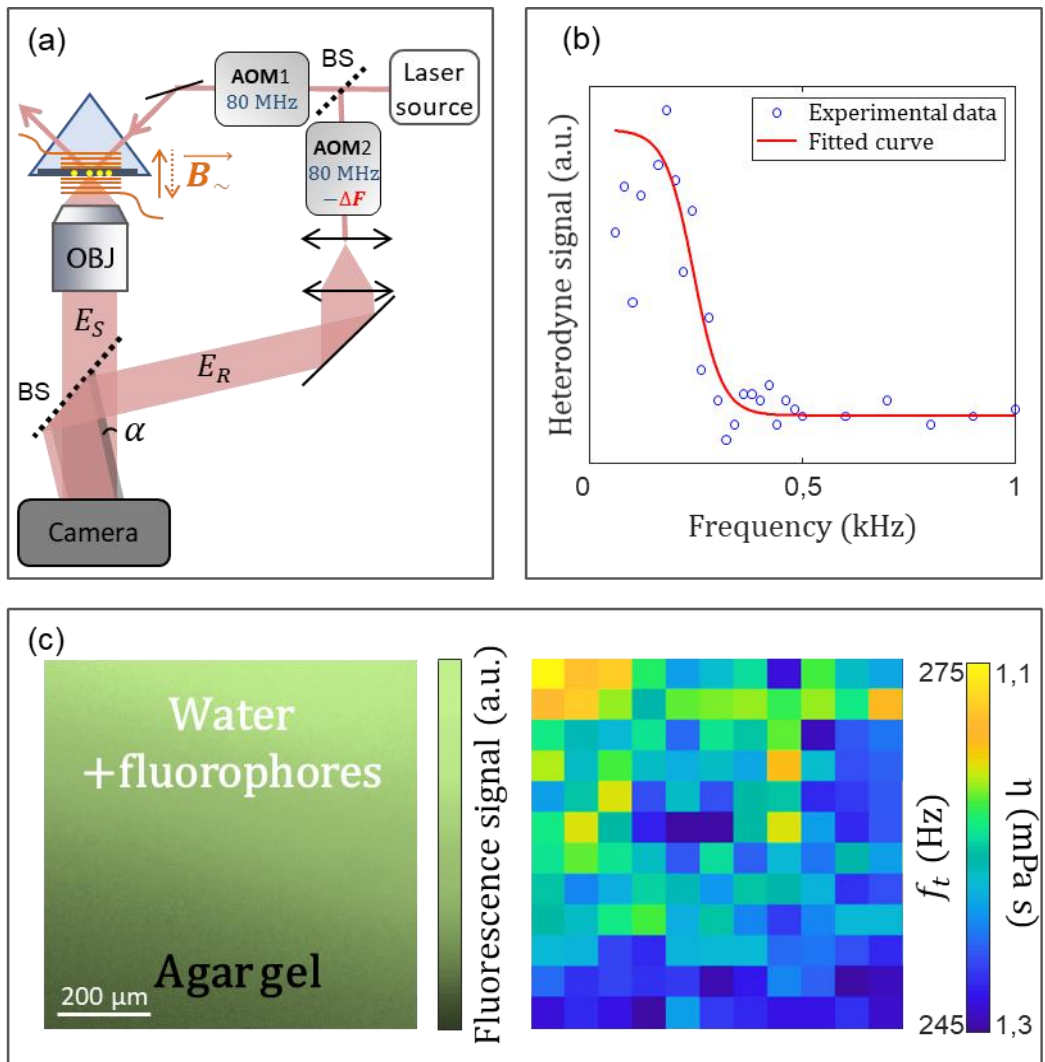

Figure 3 (a) Heterodyne holography setup: acousto-optic modulators (AOM) are added in both the object and reference arms of the holographic setup to induce a beating at $\Delta f = \frac{f_{acq}}{4} + f_i$ and selectively retrieve $f_i$ components after 4-phase demodulation. (b) Experimental spectrum and sigmoidal fit obtained for one of the pixels of (c). Similar fits conducted on each (x,y) pixels yield (c) images of the local transition frequency $f_t$ or of the quantitative viscosity (right). The observed gradient agrees with the qualitative fluorescence viscosity image (left).

frequencies $f_i$. The $1024 \times 1024$ pixels image were then resampled into $40 \times 40$ macropixels, a size corresponding to $65 \times 65$ μm² in the object space, which was chosen as a trade-off between spatial resolution and SNR. The resulting hyperspectral ($40 \times 40 \times 27$) cube of $(x, y, f_i)$ data was then used to obtain spectral fits (i.e. along $f_i$) using the sigmoidal fitting function described above (see Supplementary section for the derivation of this function). Note that the slope parameter of the sigmoid was set equal to the value of the camera integration time as being the slowest response time of the system. These fits yield transition frequencies $f_t(x, y)$ for each pixel, which are converted into a $40 \times 40$ image of the local viscosity values $\eta(x, y)$ using Eq. (1), as shown in Fig. 3c.

As shown in Fig. 3c, the fluorescence image indicates that a vertical viscosity profile is expected, with quasi-pure water at the top, and a low Agar concentration (< 5%) at the bottom of the image. This is indeed observed qualitatively in the viscosity image (right), although some evolution of the gradient is likely to occur during the 5 min total acquisition time. The [245,275] Hz range of $f_t$ values corresponds to a viscosity in the [1.1,1.3] mPa s range, close to values expected in pure water, i.e. very low Agar concentration. This underlines the sensitivity of the method, which allows to identify weak viscosity variations.

### 4. On single particle tracking

The experiments shown in Fig.3 involve relatively large macropixels (65 µm laterally), each addressing a large number (typ. thousands) of nanorods. With sufficient SNR, this volume can be easily reduced down to the size of the PSF of the microscope. Further reduction is possible provided that less than one particle is present in the volume of the PSF: calculating the center of mass of the holographic image of a rod provides its 3D position with nm-range precision. In liquids, Brownian motion ensures a stochastic exploration of the environment by the rods, and we have shown that this can provide sub-diffraction optical imaging in 3D [42]. However, in the context of viscosity measurements, the measurement time $\tau_{meas}$ must be long enough to allow rod rotation by one or more turns. During this time, the Brownian rod is free to travel laterally over a mean distance $\sqrt{D_{lat}\tau_{meas}}$, with $D_{lat}$ the lateral diffusion coefficient, which determines the size of the probed volume, i.e. the spatial resolution.

Fig. 4a shows the trajectories of a nanorod measured by superlocalization on holographic reconstructions recorded at $f_{acq} = 20$ Hz with a ×50, 0.6 NA objective, with a magnetic field $B_\sim = 3.5 \pm 0.3$ mT and $B_0 = 3.0 \pm 0.4$ mT. The volume explored by the rod during $\tau_{meas} = N/f_{acq}$ (with $N = 40$ to $1000$ the number of holograms) decreases down to $0.5$ μm³ for $N = 40$. Clearly, the SNR of the heterodyne signal also decreases with $N$. However, the transition frequency $f_t$ is well identified in measurements obtained at 9 frequencies by averaging $N = 40$ holograms in 3 glycerol solutions of respective viscosities $1.0 \pm 0.1$, $2.1 \pm 0.2$ and $3.0 \pm 0.4$ mPa s. Fits on these data yield viscosities in good agreement with these expected values, showing that $N = 40$ averages provide a reliable measurement of the viscosity. Several frequencies $f_i$ are arguably necessary to derive these values, but assuming that only one measurement is taken, which would be possible by heterodyne frequency multiplexing[44], the value $(0.5$ μm³$)^{1/3} = 800$ nm therefore indicates the best achievable resolution at this point. This corresponds to the diffraction limit $\left(\frac{1.22\lambda}{2NA} = 800 \text{ nm}\right)$ laterally, but clearly surpasses it longitudinally. With higher SNR, by e.g. increasing the optical intensity or $B$, optical superresolution by 3D rod superlocalization is therefore achievable.

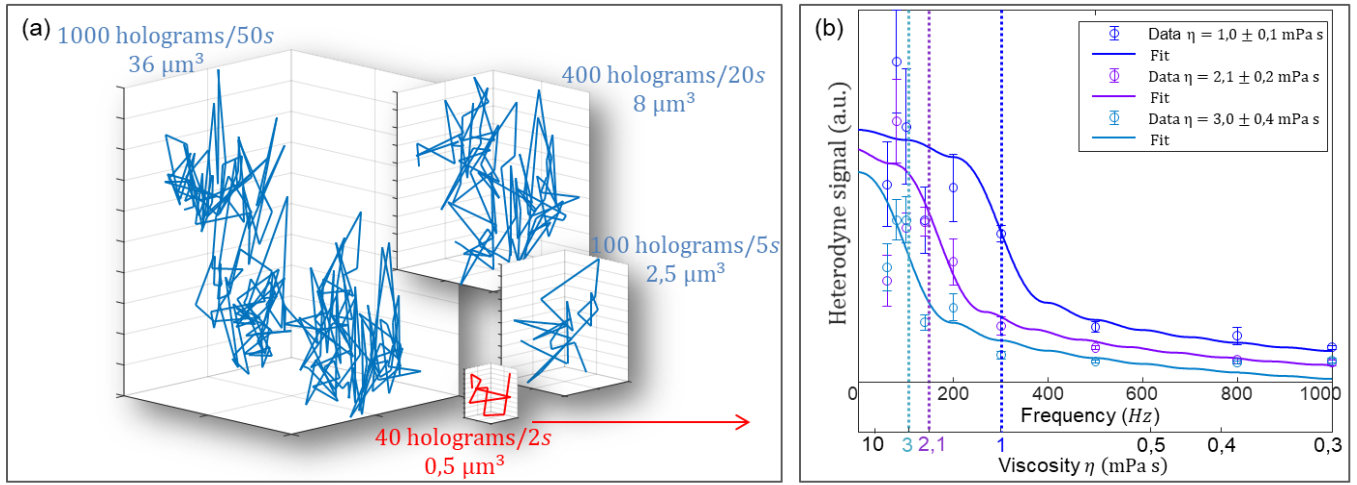

Figure 4 (a) Volumes explored by a maghemite nanorod during $N = 40$ to 1000 heterodyne holography measurements at a rate $f_{acq} = 20$ Hz. The trajectories were determined by superlocalization on 3D reconstructions. (b) Spectra of single magnetic rod rotation in water-glycerol mixtures with viscosities $1.0 \pm 0.1$, $2.1 \pm 0.2$ and $3.0 \pm 0.4$ mPa s with $B_\sim = 3.5$ mT. The fits are performed by convolution between the $sinc$ instrumental transfer function and a sigmoid function (see Supplementary), with two free parameters: the transition frequency $f_t$ and an offset $a$.

**Conclusion**

Although well tolerated and easily internalized by living cells, optically resolved maghemite needles with lengths of the order of 10 µm are not adapted to sub-cellular resolution measurements. Passive, Brownian-motion based methods with sub-100 nm gold nanorods[30] are a possible alternative since they are also biocompatible, and offer relatively large optical scattering cross sections near their plasmon resonance. However, analyzing random motion to extract viscosity values demands a fine sampling and long acquisition times which reduce the effective spatial resolution of the viscosity image. The sensitivity of the holographic detection proposed here definitely allows a drastic reduction of the probe size compared to the works which inspired it[12-14, 18]. We have shown here that the magnetically-driven rotation of non-plasmonic rods can be detected selectively at frequencies up to the MHz for $L \approx 1$ µm. The current photonic budget of the holographic detection indicates that a size reduction to a few hundreds of nm will be achievable as soon as our chemical synthesis protocol[19] allows their fabrication. Coating maghemite rods with gold might be an interesting option as it would increase their scattering cross section.

To further improve the locality and the spatial resolution, fast- and high-SNR-measurements are crucial. Increasing the amplitude of the modulated field $B_\sim$ is an efficient way to overcome rotational Brownian fluctuations. As discussed here, the transition frequency $f_t$ increases with $B_\sim^2$ and this points to high frequency measurements. Up to the kHz, high speed cameras are a solution, but short exposures call for intense illuminations which can affect the local viscosity by heating up the probes. The heterodyning scheme used here can isolate blinking nanorods well above the MHz using mW-range illumination. Since two frequencies are in principle sufficient to determine $f_t$ in media where the expected viscosity range is known, holographic multiplexing is also a possible way to improve measurement speed and hence spatial resolution. The geometry of the holographic assembly, the vertical direction of the rotation that it imposes and the presence of the prism (for dark-field illumination) complicate the use of a rotating field scheme by four identical coils[12-14, 18]. The oscillating field configuration shown in figures 2 and 3 is a good compromise between simplicity of fabrication and high enough field amplitude values. The price to pay is that the critical phenomenon allowing viscosity measurements loses accuracy: with a rotating field, a sharp cutoff between complete rotation and oscillation around a fixed direction is observed, whereas an oscillating field requires the determination of a smoother inflection point. The transition between rod oscillation and static alignment (see Supplementary) is defined with a lower precision.

Holography has the advantage of allowing 3D reconstructions. Although the PSF extension is larger in the axial direction, three-dimensional particle localization (Fig.4) can provide 3D viscosity mapping. Conversely, without localization, a pixel contains the average of signals from multiple planes (Fig.3c). There is no axial sectioning, except for the weighting of out-of-focus signals according to their pixel spread. This can be problematic in the case of thick and highly scattering samples. In order to achieve axial sectioning at excitation, a holography-OCT hybridization consisting in the use of a weakly incoherent source on the holographic setup may be a relevant strategy.

# SUPPLEMENTARY INFORMATION

## 1: Superparamagnetic rod motion

An object made of a ferromagnetic material whose size is comparable to or smaller than a ferromagnetic domain can be considered as a so called "macro-spin". When averaging the magnetic moment over an assembly of these objects or by ergodicity, in a single object over a long enough period of time, the resulting magnetic moment vanishes in the absence of an external magnetic field, and is referred to as "superparamagnetic".

For this class of objects, which includes maghemite nanorods, superparamagnetism appears under the condition $T \gg T_N$ where $T_N = \frac{E_b}{k_B} \ln \frac{\tau_0}{\tau}$, the Néel temperature, is the minimum temperature allowing the thermalization of the elementary moments and a resulting zero magnetic moment when averaged over $\tau \gg \tau_0$. $E_b$ is the energy barrier required to flip the magnetic moment and $\tau_0$ is a characteristic time of the material. The barrier energy $E_b$ is therefore negligible in this condition ($k_B T \gg E_b$) [45] and the magnetic energy then reduces to the Zeeman term $E_{p_{mag}} = -\vec{m} \cdot \vec{B}$. Its minimization leads to the alignment of the magnetic moment $\vec{m}$ along the field direction. In a superparamagnetic particle of volume $V$, far from saturation, magnetization results in a macro-spin of moment:

$$\vec{m} = \frac{\bar{\bar{\chi}} V}{\mu_0} \vec{B}$$
(2)

where the susceptibility tensor $\bar{\bar{\chi}}$, for a superparamagnetic rod with respective long and short axes susceptibilities $\chi_\parallel$ and $\chi_\perp$, reads:

$$\bar{\bar{\chi}} = \begin{pmatrix} \chi_\perp & 0 & 0 \\ 0 & \chi_\perp & 0 \\ 0 & 0 & \chi_\parallel \end{pmatrix}$$
(3)

This tensor is expressed along $\vec{u_3}$, the long axis unit vector, and $\vec{u_1}$ and $\vec{u_2}$, two arbitrary orthonormal vectors of its perpendicular plane. Out of equilibrium, equations (2) and (3) imply that the magnetic moment $\vec{m}$ of the superparamagnetic rod is given by a linear superposition of the magnetic field $\vec{B}$ components, weighed by the elements of the susceptibility tensor:

$$\vec{m} = \frac{V}{\mu_0}(\chi_\perp B_1 \vec{u_1} + \chi_\perp B_2 \vec{u_2} + \chi_\parallel B_3 \vec{u_3})$$
(4)

In the regime of low Reynolds number, where hydrodynamics is ruled by Stokes equations, one can assume a linear relationship between the particle kinematic torsor (velocity and rotation vectors) $\begin{pmatrix}\vec{U}\\\vec{\Omega}\end{pmatrix}$ and the effort torsor (resulting force and moment) $\begin{pmatrix}\vec{F}\\\vec{\tau}\end{pmatrix}$ that it undergoes:

$$\begin{pmatrix}\vec{U}\\\vec{\Omega}\end{pmatrix} = M\begin{pmatrix}\vec{F}\\\vec{\tau}\end{pmatrix}$$
(5)

Velocity, rotation, force and torque refer to a common arbitrary point of the rod, chosen in general as the hydrodynamic center. The latter, under common symmetry conditions coincides with the center of mass. $M$ is the $6 \times 6$ mobility tensor split into four $3 \times 3$ respective sub-tensors of namely translation $M^t$, rotation $M^r$ and translation-rotation coupling $M^{tr}$ and transposed $M^{rt}$ [46]:

$$M = \begin{pmatrix} M^t & M^{tr} \\ (M^{tr})^T & M^r \end{pmatrix}$$
(6)

**2: General rate equations**

From equation (5), the time-evolution equations of a particle with center position $\vec{r}$ are:

$$\frac{d\vec{r}}{dt} = \vec{U} = M^t \vec{F} + M^{tr} \vec{\tau}$$
(7)

$$\frac{d\vec{u_i}}{dt} = \vec{\Omega} \times \vec{u_i} = \left((M^{tr})^T \vec{F} + M^r \vec{\tau}\right) \times \vec{u_i}$$
(8)

In a spatially uniform magnetic field where $\vec{\nabla} \cdot \vec{B} = 0$ and $\vec{\nabla} \times \vec{B} = \vec{0}$, the magnetic force $\vec{F} = \vec{\nabla}(\vec{m} \cdot \vec{B})$ is equal to zero. Only pure rotational motion is then induced. From symmetry considerations[47], one also has $M^{tr} = M^{rt} = 0$. Moreover, $M^r$ is diagonal. Equations (7) and (8) now read:

$$\frac{d\vec{r}}{dt} = \vec{0}$$
(9)

$$\frac{d\vec{u_i}}{dt} = (M^r \vec{\tau}) \times \vec{u_i}$$
(10)

Equation (10) describes the deterministic time-evolution of the eigenaxis $\vec{u_i}$ of the superparamagnetic rod subject to a torque $\vec{\tau} = \vec{m} \times \vec{B} = \frac{\overline{\chi}V}{\mu_0}\vec{B} \times \vec{B}$. $M^r_{ii}$ matrix elements depend on viscosity $\eta$ and on the object geometry via the depolarization factors $N_{\parallel,\perp}$ (defined after equation (1)) as $M^r_\parallel = \frac{N_\parallel}{2\eta V}$ and $M^r_\perp = \frac{L^2 N_\parallel + D^2 N_\perp}{2\eta V(L^2+D^2)}$ [47].

The three vector differential equations (10) lead to 9 scalar equations when projecting each onto a reference basis $(\vec{u_x}, \vec{u_y}, \vec{u_z})$.

## 3. Component Rate Equations orthogonal oscillating and fixed components.

In the experimental scheme developed in the article, the magnetic field is defined as $\vec{B} = B_\sim cos(2\pi F_B t)\vec{u_z} + B_0 \vec{u_x}$ where $(\vec{u_x}, \vec{u_y}, \vec{u_z})$ is the laboratory referential.

Brownian diffusion effects can be introduced in the simulations by adding to (10) a stochastic term[48] and replacing deterministic parameters by their expectation values. However, it can be shown that only the direction transverse to the plane defined by the field rotation is affected by the Brownian contribution. Thus, the Brownian motion only reduces the rotation amplitude in the plane of interest, and thereby the signal amplitude, without changing the rotation frequency. Moreover, the mean direction of the major axis inevitably tends to align in the plane of the oscillating field, whatever the magnitude of the Brownian contribution. Brownian motion will therefore be neglected. We define the orthonormal moving referential $(\vec{\mathcal{u}_x}, \vec{\mathcal{u}_y}, \vec{\mathcal{u}_z})$ aligned with the magnetic field such that $\vec{\mathcal{u}_x}$ is the unitary vector of $\vec{B}$, $\vec{\mathcal{u}_y}$ is upward vertical and $\vec{\mathcal{u}_y} = \vec{\mathcal{u}_z} \times \vec{\mathcal{u}_x}$. Defining $v_{i\alpha} = \vec{\mathcal{u}_i} \cdot \vec{u_\alpha}$, $h = \frac{B_\sim}{B_0}$, where $i \in \{1,2,3\}$ and $\alpha \in \{x, y, z\}$, inserting in (9) and projecting leads to the following set of 3 differential equations for $v_{3\alpha}$, (omitting those for $v_{1\alpha}$ and $v_{2\alpha}$):

$$\frac{dv_{3x}}{dt} = 2\pi f'_t[1 + h^2 cos^2(2\pi ft)]v_{3x}(1 - v_{3x}^2) + v_{3y}\frac{2\pi hf sin(2\pi ft)}{[1 + h^2 cos^2(\omega t)]}$$

$$\frac{dv_{3y}}{dt} = -2\pi f'_t[1 + h^2 cos^2(2\pi ft)]v_{3x}^2 v_{3y} - v_{3x}\frac{2\pi hf sin(2\pi ft)}{[1 + h^2 cos^2(\omega t)]}$$

(11)

$$\frac{dv_{3z}}{dt} = -2\pi f'_t[1 + h^2 cos^2(2\pi ft)]v_{3x}^2 v_{3z}$$

where $f'_t = \frac{\mu_0}{2\pi}\Delta\chi H_x^2 M_\perp^r = \frac{1}{12\pi\mu_0}\Delta\chi B_x^2 M_\perp^r$ and $M_\perp^r = M_{11}^r = M_{22}^r$. These coefficients depend on the specific shape of the rod. In our experimental conditions the mobility matrix coefficients of the rod of length $L$ and diameter $D$ can be expressed as[48]:

$$M_\perp^r = \frac{1}{\eta}g\left(\frac{L}{D}\right) = \frac{1}{\eta}\frac{L^2 N_\parallel + D^2 N_\perp}{2(L^2 + D^2)}$$

where $N_\perp$ and $N_\parallel$ are the depolarization factors. In most cases the rods can be assimilated to a prolate spheroid whose depolarization factors can be calculated in terms of the eccentricity $e = \sqrt{1 - \left(\frac{D}{L}\right)^2}$ as[48]:

$$N_\parallel = \frac{1 - e^2}{2e^3}\left(ln\left(\frac{1 + e}{1 - e}\right) - 2e\right)$$

$$N_\perp = \frac{1 - N_\parallel}{2}$$

$$\Delta\chi = \frac{\chi^2(N_\perp - N_\parallel)}{(1 + \chi N_\parallel)(1 + \chi N_\perp)}$$

is the magnetic susceptibility anisotropy factor expressed for a rod composed of an anisotropic assembly of isotropic spherical nanoparticles of massive susceptibility $\chi$ [39]. Therefore, one can express the transition frequency as:

$$f'_t = \frac{1}{2\pi\mu_0}\frac{\Delta\chi B^2}{\eta}g\left(\frac{L}{D}\right)$$
(12)

One must note that the frequency term to be considered equivalent to the cutoff expression in the case of a rotating field should be:

$$f_t = \frac{1}{2\pi\mu_0}\frac{\Delta\chi B^2}{\eta}g\left(\frac{L}{D}\right)[1+h^2\cos^2(2\pi ft)]$$

This expression requires interpretation as it appears as a time-dependent frequency. This translates in the frequency domain as a $h^2$-dependent overmodulation of the response spectrum envelope. However, if we consider the time average of this expression, we rather define the effective transition frequency as:

$$f_t = \frac{1}{2\pi\mu_0}\frac{\Delta\chi B^2}{\eta}g\left(\frac{L}{D}\right)\left(1+\frac{h^2}{2}\right)$$
(1)

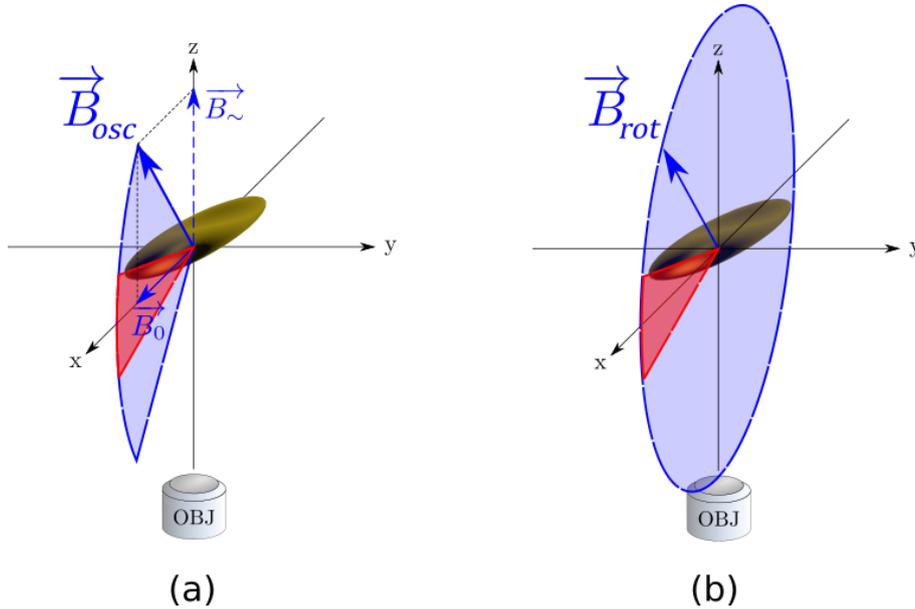

(a)                   (b)

Supplementary figure 5: excursion of a magnetic rod in a (a) oscillating and (b) rotating magnetic field. In both cases, the blue area represents the angular field covered when $f \ll f_t$. When $f > f_t$, the rod becomes unable to follow the whole angle covered by the magnetic field, and the rod only covers the angular range corresponding to the red area. With an oscillating field (a), the angle explored by the rod is thus smoothly reduced from the blue area to the red one, which is centered around the static $\vec{B_0}$ field direction. In contrast, with a rotating field (b), the rod angular exploration abruptly changes from a full rotation (blue disk) to the red angular sector which is centered around an arbitrary axis. Therefore, the rotating field (b) case defines a much sharper cut-off that illustrates a real phase transition in the rod evolution.

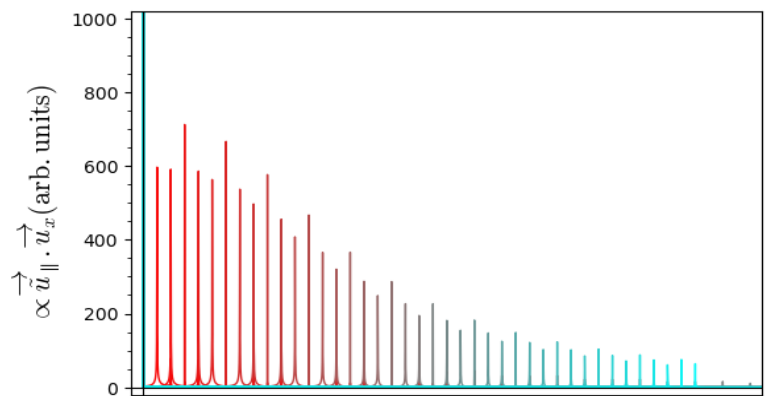
Supplementary figure 6: Frequency spectrum of the rod long axis projected on the detection plane. The envelope of the spectrum is a sigmoid, with an inflexion point near the frequency defined in equation (1).